\documentclass[aps,prd,superscriptaddress,notitlepage,floatfix,twocolumn]{revtex4-1}

\usepackage{amssymb}    
\usepackage{mathtools}
\usepackage{float}
\usepackage{subcaption}
\usepackage[colorlinks=true,
            linkcolor=red,
            urlcolor=blue,
            citecolor=green,
            bookmarks=true,
            bookmarksnumbered=true,
            breaklinks=true,
            pdfpagemode=Fullscreen,
            pdfstartview=FitBH]{hyperref}
\begin{document}

\title{Residual Symmetries Applied to Neutrino Oscillations at NO$\nu$A and T2K}

\author{Andrew D. Hanlon}
\email{ddhanlon@gmail.com}
\affiliation{Department of Physics and Astronomy, Michigan State University, East Lansing, MI 48824, USA}


\author{Wayne W. Repko}
\email{repko@pa.msu.edu}
\affiliation{Department of Physics and Astronomy, Michigan State University, East Lansing, MI 48824, USA}

\author{Duane A. Dicus}
\email{dicus@physics.utexas.edu}
\affiliation{Department of Physics and Center for Particles and Fields, University of Texas, Austin, TX 78712, USA}

\date{\today}

\begin{abstract}
The results previously obtained from the model-independent application of a generalized hidden horizontal
$\mathbb{Z}_2$ symmetry to the neutrino mass matrix are updated using the latest global fits for the neutrino
oscillation parameters. The resulting prediction for the Dirac $CP$ phase $\delta_D$ is in agreement with recent
results from T2K. The distribution for the Jarlskog invariant $J_\nu$ has become sharper and appears to be
approaching a particular region. The approximate effects of matter on long baseline
neutrino experiments are explored, and it is shown how the weak interactions between the neutrinos and the
particles that make up the Earth can help to determine the mass hierarchy. A similar strategy is employed
to show how NO$\nu$A and T2K could determine the octant of $\theta_a (\equiv \theta_{23})$.
Finally, the exact effects of matter are obtained numerically in order to make comparisons with the form
of the approximate solutions. From this analysis there emerges some interesting features of the effective
mass eigenvalues.
\end{abstract}

\maketitle

\section{Introduction}

\vspace{10pt}

Although there has been significant progress in neutrino physics from oscillation experiments, there remains
much work to be done. The reactor angle $\theta_r (\equiv \theta_{13})$ has now been measured to greater accuracy than ever before, and the solar
angle $\theta_s (\equiv \theta_{12})$ has been known for some time now. But, the question of the octant for the atmospheric angle
($\theta_a > \frac{\pi}{2}$ or $\theta_a < \frac{\pi}{2}$), or whether this angle is maximal ($\theta_a = \frac{\pi}{2}$),
has yet to be answered. Determination of the Dirac $CP$ phase has been improved.
Recent results from T2K exclude at $90\%$ C.L. $\delta_D \in [34.2^\circ, 144^\circ]$ for normal hierarchy (NH) and 
$\delta_D \in [-180^\circ, -174.6^\circ] \cup [-7.2^\circ, 180^\circ]$ for inverted hierarchy (IH) \cite{Abe:2013hdq}.
Finally, the absolute value of the mass squared differences have been carefully measured,
but the mass hierarchy is still undetermined (i.e. $m_3 \gg m_2 > m_1$, or $m_2 > m_1 \gg m_3$). Each of these questions will be discussed in this work.

\vspace{12pt}

From the improvements in recent global analyses \cite{Capozzi:2013csa,Tortola:2012te,GonzalezGarcia:2012sz} it is possible to make
more accurate predictions for the distributions of some of the aforementioned parameters of interest.
Specifically, each of the residual symmetries, $\mathbb{Z}_2^s$ and $\overline{\mathbb{Z}}^s_2$,
can be used to derive a model-independent equation for $\delta_D$
(one for each symmetry) \cite{PhysRevLett.108.041801,Hanlon:2013ska}. Then using the newly available global
fits of the neutrino oscillation parameters in
\cite{Capozzi:2013csa}, likelihood distributions for $\delta_D$, the Jarlskog invariant \cite{PhysRevLett.55.1039}, and $\theta_a$ are obtained.

\vspace{12pt}

Using the PMNS mixing matrix, an expression for the probability of a neutrino originally of flavor $\alpha$ to be detected
as a neutrino of flavor $\beta$, $P(\nu_\alpha \to \nu_\beta)$, is presented (which is a standard result found
in many review papers on neutrino physics \cite{Kayser:2005cd,Kayser:2008ev,Kayser:2008rd}).
Then, using the approximation from \cite{Agarwalla:2013tza} it is shown how the earth's matter
affects the neutrino beam in long baseline experiments. This is done by replacing the oscillation parameters
with effective values that depend on the energy of the neutrinos, the baseline length, and the density of the matter.

\vspace{12pt}

In this paper, a focus is made on the NO$\nu$A and T2K experiments. Both of these experiments measure the
appearance of $\nu_e$'s ($\overline{\nu}_e$'s) from a $\nu_\mu$ ($\overline{\nu}_\mu$) beam.
The probability for this appearance is plotted as a function of energy using the best fits for the oscillation parameters
in \cite{Capozzi:2013csa}. The effects of matter are taken into account using the average matter density
along the baseline for the two experiments. This is justified by the fact that there does not appear to be a significant
effect due to the variation of the matter density.
A comparison is made for this probability with and without $CP$-violation in an attempt
to observe the sensitivity of NO$\nu$A and T2K to measurements of $\delta_D$. We have also plotted
$P(\nu_\mu \to \nu_e)$ vs. $P(\overline{\nu}_\mu \to \overline{\nu}_e)$ which shows that it may be possible
for these experiments to determine the neutrino mass hierarchy for some values of the $CP$ phase
as discussed in \cite{Patterson:2012zs,Agarwalla:2013ju}.

\vspace{12pt}

The update of the analysis of \cite{PhysRevD.86.013012} given in \cite{Capozzi:2013csa} gives closer
agreement on $\theta_a$ with the other two major global analyses \cite{GonzalezGarcia:2012sz,Tortola:2012te}.
This shows that $\theta_a$ is closer to being maximal than originally believed and only excludes
the possibility of it being maximal by about $1\sigma$ for inverted hierarchy. But, it
is clear that the analyses do not agree upon which octant is favored.
Fortunately, the plots of $P(\nu_\mu \to \nu_e)$ vs. $P(\overline{\nu}_\mu \to \overline{\nu}_e)$ may
also serve to determine the octant of $\theta_a$ \cite{Patterson:2012zs,Agarwalla:2013ju}.

\vspace{12pt}

This work concludes with a digression into the effective mixing angles and masses in matter.
The solar resonance, first described by the MSW effect \cite{Wolfenstein:1977ue,Mikheev:1986wj,Smirnov:2004zv}, and the atmospheric resonance
are readily observed.

\section{Distribution of $\delta_D$, $J_\nu$, and $\theta_a$}

The equations for $\delta_D$, in terms of the neutrino mixing angles, based on residual symmetries
are given by \cite{PhysRevLett.108.041801,Hanlon:2013ska},
\begin{subequations} \label{eq:cD}
  \begin{align}
    \cos \delta_D = \frac{(s_s^2 - c_s^2 s_r^2)(s_a^2 - c_a^2)}{4 c_a s_a c_s s_s s_r} \label{eq:cD_a},\\
    \cos \delta_D = \frac{(s_s^2 s_r^2 - c_s^2)(s_a^2 - c_a^2)}{4 c_a s_a c_s s_s s_r} \label{eq:cD_b},
  \end{align}
\end{subequations}
for $\mathbb{Z}_2^s$ and $\overline{\mathbb{Z}}_2^s$ respectively, where $s_i \equiv \sin \theta_i$, and
$c_i \equiv \cos \theta_i$. The latest global fits for the neutrino
oscillation parameters from \cite{Capozzi:2013csa} are shown in Table \ref{tab:mixing_angles}.

\begin{table}[H]
\centering
  \begin{tabular}{| c || c | c |}
    \hline
    & & \\[-0.7em]
    Parameter & Best fit & 1$\sigma$ range \\[0.75ex]
    \hline
    & &\\[-.7em]
    $\sin^2 \theta_s/10^{-1}$ (NH or IH) & 3.08 & 2.91-3.25\\[.75ex]
    \hline
    & &\\[-.7em]
    $\sin^2 \theta_r/10^{-2}$ (NH) & 2.34 & 2.16-2.56\\[.75ex]
    \hline
    & &\\[-.7em]
    $\sin^2 \theta_r/10^{-2}$ (IH) & 2.39 & 2.18-2.60\\[.75ex]
    \hline
    & &\\[-.7em]
    $\sin^2 \theta_a/10^{-1}$ (NH) & 4.25 & 3.98-4.54\\[.75ex]
    \hline
    & &\\[-.7em]
    $\sin^2 \theta_a/10^{-1}$ (IH) & 4.37, 5.82* & 4.08-4.96 $\oplus$ 5.31-6.10\\[.75ex]
    \hline
    & &\\[-.7em]
    $\delta_D / \pi$ (NH) & 1.39 & 1.12-1.72\\[.75ex]
    \hline
    & &\\[-.7em]
    $\delta_D / \pi$ (IH) & 1.35 & 0.96-1.59\\[.75ex]
    \hline
    & &\\[-0.7em]
    $m_{21}^2/10^{-5} eV^2$ (NH or IH) & 7.54 & 7.32-7.80\\[.75ex]
    \hline
    & &\\[-0.7em]
    $m_{31}^2/10^{-3} eV^2$ (NH) & 2.48 & 2.42-2.56 \\[.75ex]
    \hline
    & &\\[-0.7em]
    $|m_{31}^2|/10^{-3} eV^2$ (IH) & 2.36 & 2.29-2.43 \\[.75ex]
    \hline
  \end{tabular}
  \caption{Global fits for neutrino oscillation parameters from \cite{Capozzi:2013csa}.
           *Represents a local minimum at approximately $0.42\sigma$ for $\chi^2$.}
  \label{tab:mixing_angles}
\end{table}

\begin{figure}[H]
  \centering
    \begin{subfigure}{0.5\textwidth}
      \centering
      \includegraphics[width=\textwidth]{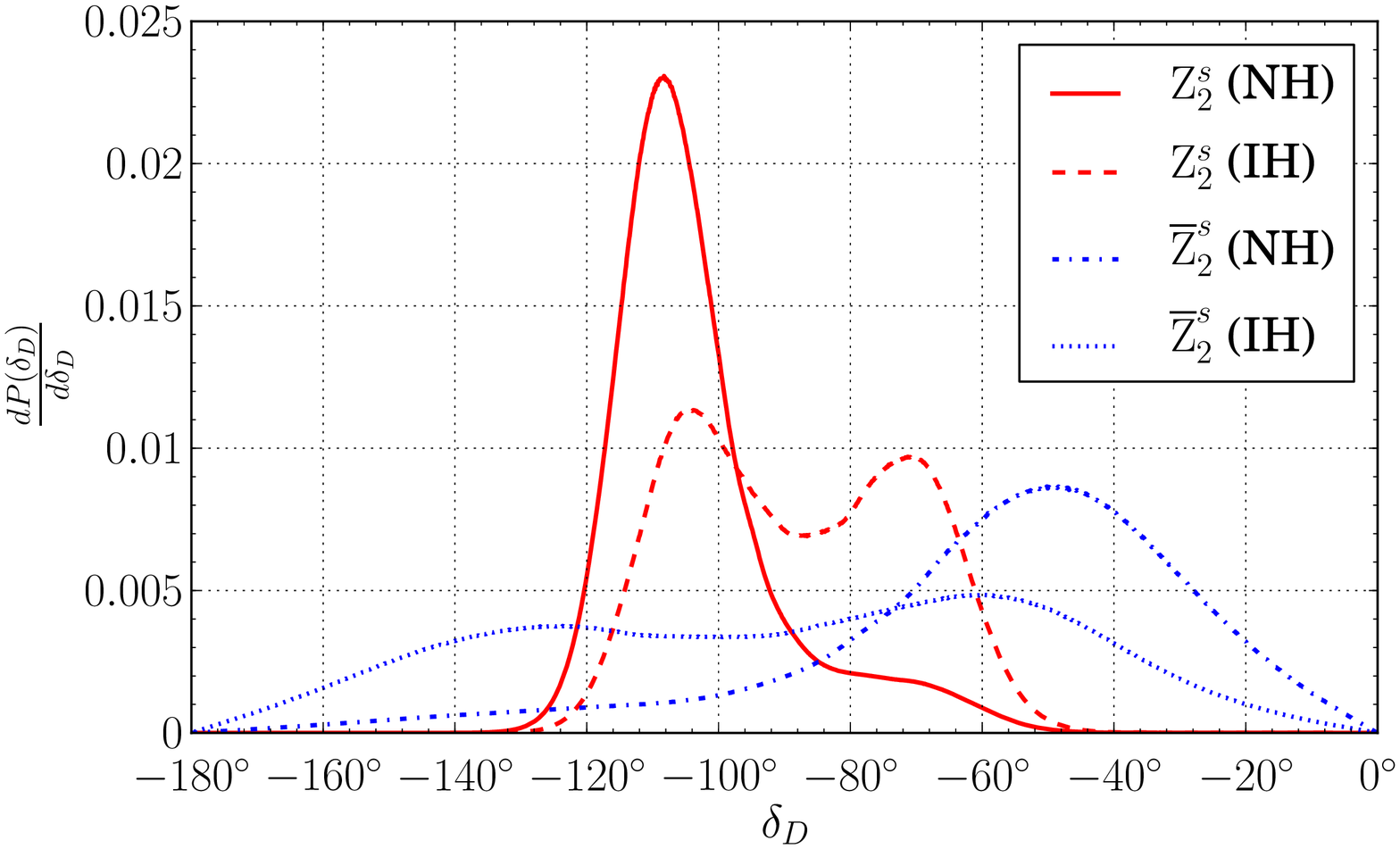}
      \caption{}
      \label{fig:deltaD}
    \end{subfigure}
    \begin{subfigure}{0.5\textwidth}
      \centering
      \includegraphics[width=\textwidth]{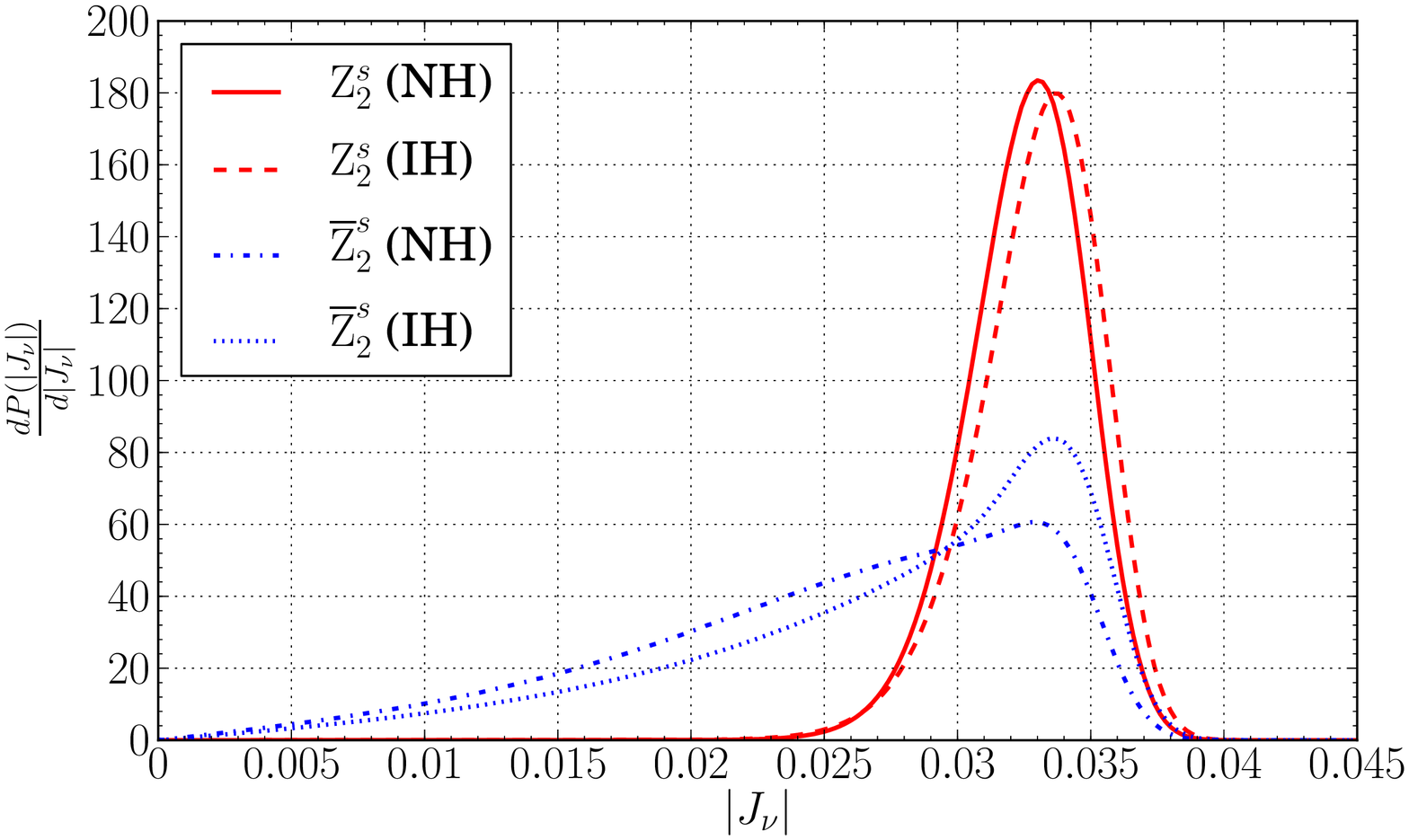}
      \caption{}
      \label{fig:jarlskog}
    \end{subfigure}
    \begin{subfigure}{0.5\textwidth}
      \centering
      \includegraphics[width=\textwidth]{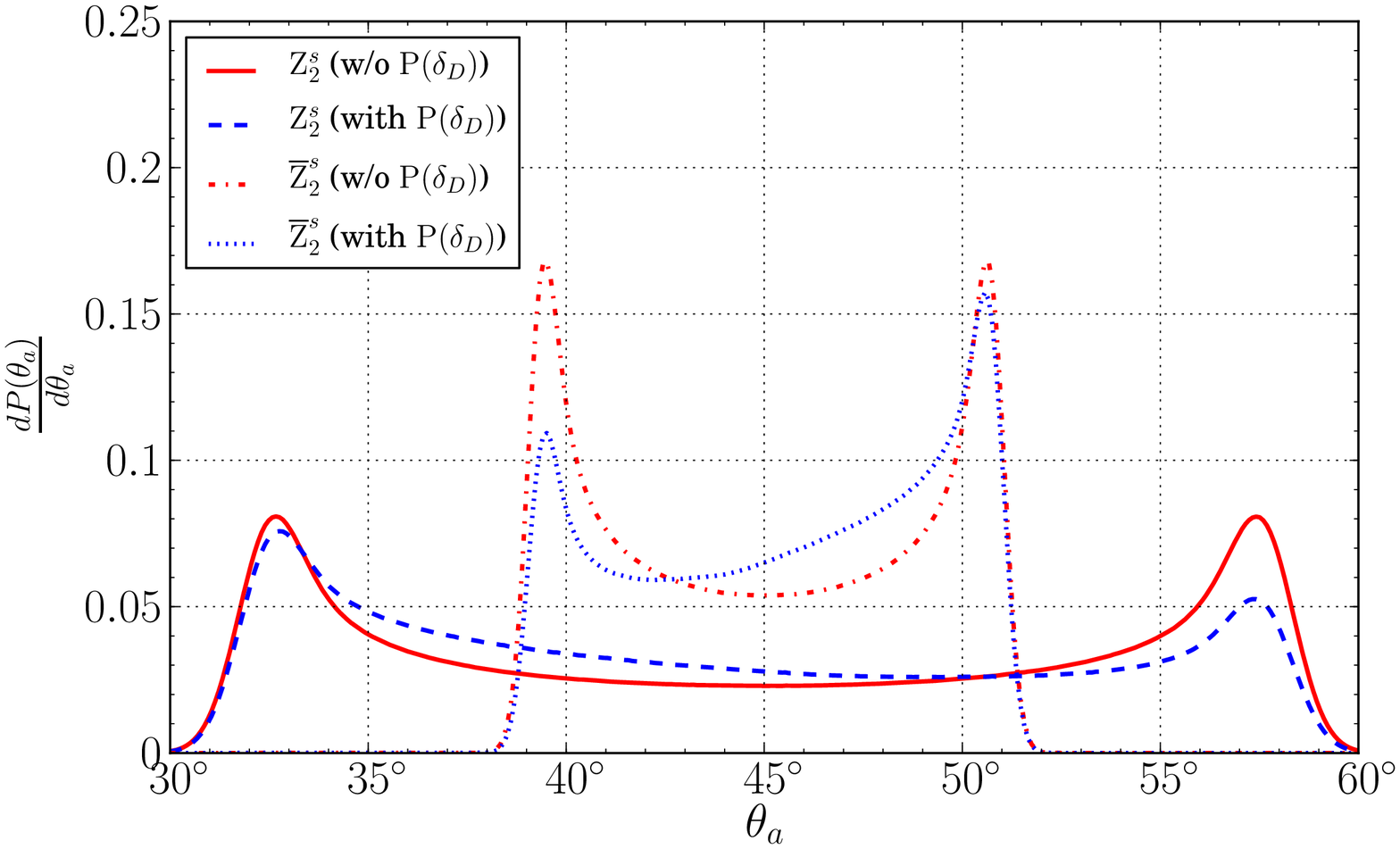}
      \caption{}
      \label{fig:thetaA}
    \end{subfigure}
    \caption{Predicted distributions for (a) $\delta_D$, (b) $J_\nu$, and (c) $\theta_a$ (NH) using the global analysis in \cite{Capozzi:2013csa}.}
  \label{fig:distributions}
\end{figure}

From this we can obtain a distribution for $\cos \delta_D$ following the procedure in \cite{Hanlon:2013ska,PhysRevLett.108.041801} by using,
\begin{equation} \label{eq:originalDist}
  \frac{d P(\cos \delta_D)}{d \cos \delta_D} = \int \delta^p_D \mathbb{P}(s_a^2) \mathbb{P}(s_s^2) \mathbb{P}(s_r^2) \mathrm{d}s_a^2 \mathrm{d}s_s^2 \mathrm{d}s_r^2,
\end{equation}
where $\delta^p_D \equiv \delta(\cos{\delta_D} - \overline{c}_D)$, the $\mathbb{P}$'s are proportional to $\exp(- \chi^2 / 2)$,
 and $\overline{c}_D \equiv$ RHS of (\ref{eq:cD}). 
Because it is preferable to get a distribution with respect to $\delta_D$ rather than $\cos \delta_D$ we use,
\begin{equation} \label{eq:changeVar}
  \frac{d P(\delta_D)}{d \delta_D} = |s_D| \frac{d P(c_D)}{d c_D},
\end{equation}
where $c_D \equiv \cos \delta_D$ and $s_D \equiv \sin \delta_D$. Since this is a numerical integral, the delta function
cannot be used as it is normally defined (unless integrated out of the equation prior to the numerical calculation).
The integral was evaluated using a Monte Carlo algorithm and
the results are shown in Fig. \ref{fig:deltaD}. The domain of $\delta_D$ in (\ref{eq:changeVar}) is $[-180^\circ,0^\circ]$, but
the distributions in Fig. \ref{fig:deltaD} can be reflected about
$\delta_D = 0^\circ$ to account for
the full interval $[-180^\circ,180^\circ]$. Therefore, these distributions have been normalized to $\frac{1}{2}$ over the domain shown in the figures. This means that
each of the residual symmetries will have two peak predictions for the $CP$ phase (equidistant from $0^\circ$). The IH $\chi^2$ curve for $\theta_a$
in \cite{Capozzi:2013csa} is closer to being symmetric about $\sin^2 \theta_a = 0.5$. This is very prevalent in the results shown in
Fig. \ref{fig:deltaD} given that the IH plots are close to being symmetric about $\delta_D = -90^\circ$. But, since the NH global fit
favors the lower octant for $\theta_a$ by at least $2 \sigma$ \cite{Capozzi:2013csa} the predicted distributions for NH tend to prefer one
side of $\delta_D = -90^\circ$. But in both cases
the results for $\mathbb{Z}^s_2$ are in agreement with the best fit value of $\delta_D = -90^\circ$ from T2K's latest results  \cite{Abe:2013hdq}.

The same method is applied to the Jarlskog invariant
$J_\nu \equiv c_a s_a c_s s_s c_r^2 s_r s_D$ \cite{PhysRevLett.55.1039}, that is,
\begin{equation} \label{eq:jarlsDist}
  \frac{d P(J_\nu)}{d J_\nu} = \int \delta^p_{J_\nu} \mathbb{P}(s_a^2) \mathbb{P}(s_s^2) \mathbb{P}(s_r^2) \mathrm{d}s_a^2 \mathrm{d}s_s^2 \mathrm{d}s_r^2,
\end{equation}
with, $\delta^p_{J_\nu} \equiv \delta(J_\nu - c_a s_a c_s s_s c_r^2 s_r s_D)$.
This distribution is shown in Fig. \ref{fig:jarlskog}. When calculating these distributions, $\delta_D$ is taken to be
in the interval $[0,180^\circ]$ and is even about the vertical axis to extend $\delta_D$
to include $[-180^\circ, 0^\circ]$.
To account for this, the figures are labeled for the distribution of $|J_\nu|$, and they can therefore be normalized to one.
As compared with our previous results in \cite{Hanlon:2013ska}, $\overline{\mathbb{Z}}^s_2$ is beginning to favor the 
region that $\mathbb{Z}^s_2$ prefers. Also, the region predicted by $\mathbb{Z}^s_2$ has become slightly narrower and it now excludes
$|J_\nu| < 0.024$.

Finally, this method is again applied similarly to $\theta_a$ by first using (\ref{eq:cD}) to solve for $\tan 2 \theta_a$,
\begin{subequations} \label{eq:thetaA}
  \begin{align}
    \tan 2 \theta_a = \frac{c_s^2 s_r^2 - s_s^2}{2 c_s s_s s_r \cos \delta_D} \label{eq:thetaA_a},\\
    \tan 2 \theta_a = \frac{c_s^2 - s_s^2 s_r^2}{2 c_s s_s s_r \cos \delta_D} \label{eq:thetaA_b},
  \end{align}
\end{subequations}
for $\mathbb{Z}_2^s$ and $\overline{\mathbb{Z}}_2^s$ respectively. Then we have,
\begin{equation} \label{eq:thetaAdist}
  \frac{d P(\tan 2 \theta_a)}{d \tan 2 \theta_a} = \int \delta^p_{\theta_a} \mathbb{P}(s_s^2) \mathbb{P}(s_r^2) \mathbb{P}(\delta_D) \mathrm{d}s_s^2 \mathrm{d}s_r^2 \mathrm{d}\delta_D,
\end{equation}
with $\delta^p_{\theta_a} \equiv \delta(\tan 2 \theta_a - \overline{t}_{\theta_a})$, where $\overline{t}_{\theta_a} \equiv$ RHS of (\ref{eq:thetaA}).
To get a distribution for $\theta_a$ we use,
\begin{equation}
  \frac{d P(\theta_a)}{d \theta_a} = 2 \sec^2 (2 \theta_a) \frac{d P(\tan 2 \theta_a)}{d \tan 2 \theta_a}.
\end{equation}
The distribution is shown in Fig. \ref{fig:thetaA}, where plots are made with and without using the prior on
$\delta_D$ from \cite{Capozzi:2013csa}. When no prior on $\delta_D$ is used, $\mathbb{P}(\delta_D)$ becomes evenly distributed in $[0, 2 \pi)$.
As previously discussed in \cite{Hanlon:2013ska}, $\theta_a$ is symmetric about $\theta_a = 45^\circ$ when there is no prior on $\delta_D$.
In addition, the distributions using the prior on $\delta_D$ have also become more symmetric, as a result of the $\chi^2$ for $\cos \delta_D$ also
having become more symmetric about zero.

\section{$\nu_{\mu}$ to $\nu_{e}$ oscillation}

Now that we have a distribution for all the neutrino oscillation parameters,
an attempt can be made to predict the results
of an experiment measuring the number of $\nu_\mu$'s that oscillate into $\nu_e$'s over some distance. 
First, the expression for this probability, $P(\nu_\mu \to \nu_e)$, must be found. Denoting the weak
eigenstates of the neutrino by $\mid \nu_{\alpha} \rangle$, and the neutrino mass eigenstates by $\mid \nu_i \rangle$, then
\begin{equation}
  (U_{PMNS})_{\alpha j} \equiv \langle \nu_\alpha \mid \nu_j \rangle,
\end{equation}
defines the PMNS mixing matrix, $U_{PMNS}$. The standard parametrization is given by \cite{Agarwalla:2013tza},
\begin{equation}
  U_{PMNS} = U \mathcal{P},
\end{equation}
where
\begin{equation}
   U = 
    \begin{pmatrix}
      c_s c_r & s_s c_r & s_r e^{-i \delta_D} \\
      - s_s c_a - c_s s_a s_r e^{i \delta_D} & c_s c_a - s_s s_a s_r e^{i \delta_D} & s_a c_r \\
      s_s s_a - c_s c_a s_r e^{i \delta_D} & - c_s s_a - s_s c_a s_r e^{i \delta_D} & c_a c_r
    \end{pmatrix},
\end{equation}
\begin{equation}
  \mathcal{P} = diag(1,e^{i \alpha_{21}/2},e^{i \alpha_{31}/2}).
\end{equation}
From \cite{Kayser:2005cd},
\begin{equation}
  Amp(\nu_\alpha \to \nu_\beta) = \sum_i U^*_{\alpha i} e^{-i m_i^2 \frac{L}{2 E}} U_{\beta i},
\end{equation}
which leads to,
\begin{multline}
  \hspace{-20pt} P \big(\overset{(-)}{\nu_\alpha} \to \overset{(-)}{\nu_\beta} \big) = \delta_{\alpha \beta}
    - 4 \sum_{i > j} \mathfrak{R}(U^*_{\alpha i} U_{\beta i} U_{\alpha j} U^*_{\beta j}) \sin^2 \Big(\Delta m^2_{i j} \frac{L}{4 E} \Big) \\
    \stackrel{+}{ \left( - \right)} 2 \sum_{i > j} \mathfrak{I}(U^*_{\alpha i} U_{\beta i} U_{\alpha j} U^*_{\beta j}) \sin \Big( \Delta m^2_{i j} \frac{L}{2 E} \Big)
\end{multline}
where $\Delta m^2_{ij} \equiv m^2_i - m^2_j$, $m_i$ is the $i$th mass eigenvalue,
$L$ is the distance propagated by the neutrino, and $E$ is the energy of the neutrino.
Notice that this probability does not depend on the Majorana phases, and therefore a discussion on these phases will not be pursued here. 

Making the following definition \cite{Agarwalla:2013tza},
\begin{equation}
  \Delta_{i j} \equiv \frac{\Delta m^2_{i j}}{2 E} L,
\end{equation}
and noting that, $\Delta_{32} = \Delta_{31} - \Delta_{21}$, then
\begin{multline} \label{eq:prob_mu_to_e}
  P \big(\overset{(-)}{\nu_\mu} \to \overset{(-)}{\nu_e} \big) =
    4 s_s^2 c_r^2 \big( s_s^2 s_r^2 s_a^2 + c_s^2 c_a^2 - 2 c_s c_a s_s s_r s_a c_D \big) \\
    \times \sin^2 \Big( \frac{\Delta_{21}}{2} \Big)
    + 4 s_r^2 s_a^2 c_r^2 \sin^2 \Big( \frac{\Delta_{31}}{2} \Big)
    + 2 s_s s_r c_r^2 s_a \big( c_s c_a c_D - \\
    s_s s_r s_a \big) 
    \Big[4 \sin^2 \Big(\frac{\Delta_{21}}{2} \Big) \sin^2 \Big(\frac{\Delta_{31}}{2} \Big)
    + \sin \big(\Delta_{21} \big) \sin \big(\Delta_{31} \big) \Big] \\
    \overset{+}{(-)} 4 J_\nu \Big[ \sin^2 \Big(\frac{\Delta_{21}}{2} \Big) \sin \big(\Delta_{31} \big) 
    - \sin^2 \Big( \frac{\Delta_{31}}{2} \Big) \sin \big(\Delta_{21} \big) \Big].
\end{multline}
The last term includes the Jarlskog invariant \cite{PhysRevLett.55.1039} defined above.

\subsection{Matter Effects}
\label{sec:matter}

As electron neutrinos propagate through the earth, they can interact with electrons via $W$-exchange. In addition,
all three neutrino flavors can interact with electrons, protons, or neutrons via $Z$-exchange. Assuming
electrically neutral matter, the $Z$-exchange between the neutrinos and protons will cancel exactly
with the $Z$-exchange between the neutrinos and electrons \cite{Kayser:2005cd}. The contribution from $Z$-exchange can be dropped, because it
only adds a multiple of the identity matrix to the Hamiltonian \cite{Agarwalla:2013tza}. Then, under the assumption that $E \ll M_W$, 
the effect of $W$-exchange can be accounted for by modifying the Hamiltonian for
neutrinos \cite{Kimura:2002wd},
\begin{equation} \label{eq:hamiltonian}
  H = \frac{1}{2 E} U
    \begin{pmatrix}
      0 & 0 & 0 \\
      0 & \Delta m^2_{21} & 0 \\
      0 & 0 & \Delta m^2_{31}
    \end{pmatrix}
    U^\dagger
    + \frac{1}{2 E}
    \begin{pmatrix}
      a & 0 & 0 \\
      0 & 0 & 0 \\
      0 & 0 & 0
    \end{pmatrix},
\end{equation}
where $a \equiv 2 \sqrt{2} G_F N_e E$, and $N_e$ is the density of electrons. For anti-neutrinos, the Hamiltonian
is simply the complex conjugate of (\ref{eq:hamiltonian}) with $a \to -a$.

One way to proceed is to diagonalize the Hamiltonian exactly, which has been done analytically
\cite{Zaglauer1988,Kimura:2002wd,Kimura:2002hb}. 
However, this doesn't give much physical insight into the effects of matter on neutrino oscillations. Approximations
in which the mixing angles and mass eigenvalues are replaced by effective values don't modify any of the equations,
and therefore it becomes clear how matter affects neutrinos.
A number of approximation schemes have been developed
\cite{Arafune:1997hd,Freund:2001pn,Peres:2003wd,Akhmedov:2004ny,Akhmedov:2004rq,Akhmedov:2008nq,Cervera:2000kp}.
One of the most commonly used of these are the equations derived in \cite{Cervera:2000kp}. But, due to the
large value of $\theta_r$ measured at Daya Bay \cite{Dwyer:2013wqa}, the approximation in
\cite{Cervera:2000kp} begins to fail as is shown in \cite{Agarwalla:2013tza}.
In the approximation that is
used here, the form of (\ref{eq:prob_mu_to_e}) can be used with the following
modifications \cite{Agarwalla:2013tza}:
\begin{multline}
  \label{eq:eff_angles}
  \hspace{60pt} \theta_{s} \to \theta^\prime_s,
  \hspace{20pt} \theta_{r} \to \theta^\prime_r, \\
  \Delta m^2_{21} \to \lambda_2 - \lambda_1,
  \hspace{20pt} \Delta m^2_{31} \to \lambda_3 - \lambda_1,
  \hspace{10pt}
\end{multline}
with,
\begin{subequations} \label{eq:mixing_angles_approx}
\begin{gather}
\tan (2 \theta^\prime_{s}) = \frac{(\Delta m^2_{21} / c^2_r) \sin (2 \theta_s) }{(\Delta m^2_{21} / c^2_r) \cos (2 \theta_s) - a}, \label{eq:theta_s_approx} \\
\tan (2 \theta^\prime_{r}) = \frac{(\Delta m^2_{31} - \Delta m^2_{21} s^2_{s}) \sin (2 \theta_{r})}{(\Delta m^2_{31} - \Delta m^2_{21} s^2_{s}) \cos (2 \theta_{r}) - a}, \label{eq:theta_r_approx} \\
\hspace{-25pt} \lambda^\prime_\pm \equiv \frac{(\Delta m^2_{21} + a c^2_{r})
    \pm \sqrt{(\Delta m^2_{21} - a c^2_r)^2 + 4 a c^2_r s^2_s \Delta m^2_{21}}}{2}, \\
\hspace{-25pt} \lambda^{\prime \prime}_\pm \equiv \frac{\lambda + (\Delta m^2_{31} + a s^2_{r})
    \pm \sqrt{[\lambda - (\Delta m^2_{31} + a s^2_r)]^2 + 4 a^2 s c^2_r s^2_r}}{2}.
\end{gather}
\end{subequations}
Where, for neutrinos let,
\begin{multline}
 \hspace{40pt} \lambda \equiv \lambda^\prime_+,
 \hspace{20pt} s \equiv s^{\prime 2}_s,
 \hspace{20pt} \lambda_1 \approx \lambda^{'}_{-}, \\
               \lambda_2 \approx \lambda^{''}_{\mp},
 \hspace{25pt} \lambda_3 \approx \lambda^{''}_{\pm},
 \hspace{45pt}
\end{multline}
and for anti-neutrinos let,
\begin{multline}
    \hspace{30pt} \lambda \equiv \lambda^\prime_-,
    \hspace{20pt} s \equiv c^{\prime 2}_s,
    \hspace{20pt} a \to -a, \\
    \lambda_1 \approx \lambda^{''}_{\mp},
    \hspace{20pt} \lambda_2 \approx \lambda^{'}_{+},
    \hspace{20pt} \lambda_3 \approx \lambda^{''}_{\pm},
    \hspace{25pt}
\end{multline}
with the upper sign for normal hierarchy and the lower sign for inverted hierarchy.

It's helpful to show $a$ and $\Delta_{i j}$ in
conventional units. Following \cite{Agarwalla:2013tza},
\begin{subequations}
  \begin{gather}
    \Delta_{i j} = 2.534 \bigg( \frac{\Delta m^2_{i j}}{[eV^2]} \bigg) \bigg(\frac{[GeV]}{E} \bigg)
      \bigg ( \frac{L}{[km]} \bigg) \\
    a = (7.63 \times 10^{-5} [eV^2]) \bigg( \frac{\rho}{[g / cm^3]} \bigg) \bigg(\frac{E}{[GeV]} \bigg). 
  \end{gather}
\end{subequations}

\subsection{NO$\nu$A and T2K}
NO$\nu$A is a long-baseline neutrino oscillation experiment located in northern Minnesota. It has a baseline
length of $810$ $km$, has an average matter density of $2.8$ $g/cm^3$ along this baseline, and a peak neutrino
energy around $2$ $GeV$ \cite{Patterson:2012zs}.
T2K is another neutrino oscillation experiment with similar goals to that of NO$\nu$A. Its baseline length is
$295$ $km$, has an average matter density of $2.6$ $g/cm^3$, and the neutrino beam energy peaks around
$0.6$ $GeV$ \cite{Hagiwara:2011kw}.

\begin{figure}[H]
  \begin{subfigure}{0.5\textwidth}
    \centering
    \includegraphics[width=\textwidth]{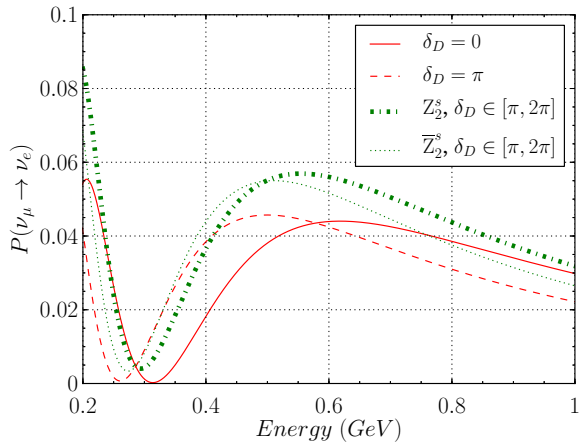}
    \caption{}
    \label{fig:oscillation_NH_T2K_fogli}
  \end{subfigure}
  \caption{$P(\nu_\mu \to \nu_e)$ as a function of energy at (a) NO$\nu$A and
           (b) T2K,
           using the best fits from \cite{Capozzi:2013csa} with normal hierarchy.}
  \label{fig:oscillation_NOvA}
\end{figure}

With the use of the effective mixing angles derived in the previous section, the probability of the appearance of a
$\nu_e$ ($\overline{\nu}_e$) from a $\nu_\mu$ ($\overline{\nu}_\mu$) beam can be determined for any matter density.
Using the length and matter density
for the two experiments in question, plots of these probabilities are shown in Fig.
\ref{fig:oscillation_NOvA} as a function of energy.

It's not entirely apparent that the approximation \cite{Agarwalla:2013tza} is valid for different values of the $CP$ phase or the
vacuum mixing angles, therefore
a comparison is made between this approximation and the exact results in the Appendix. In this comparison, the exact
results are found by numerically diagonalizing the Hamiltonian.
As it turns out, the approximation is very good
for the energies and densities considered here.
\begin{figure}[H]
  \centering
  \begin{subfigure}{0.5\textwidth}
    \centering
    \includegraphics[width=\textwidth]{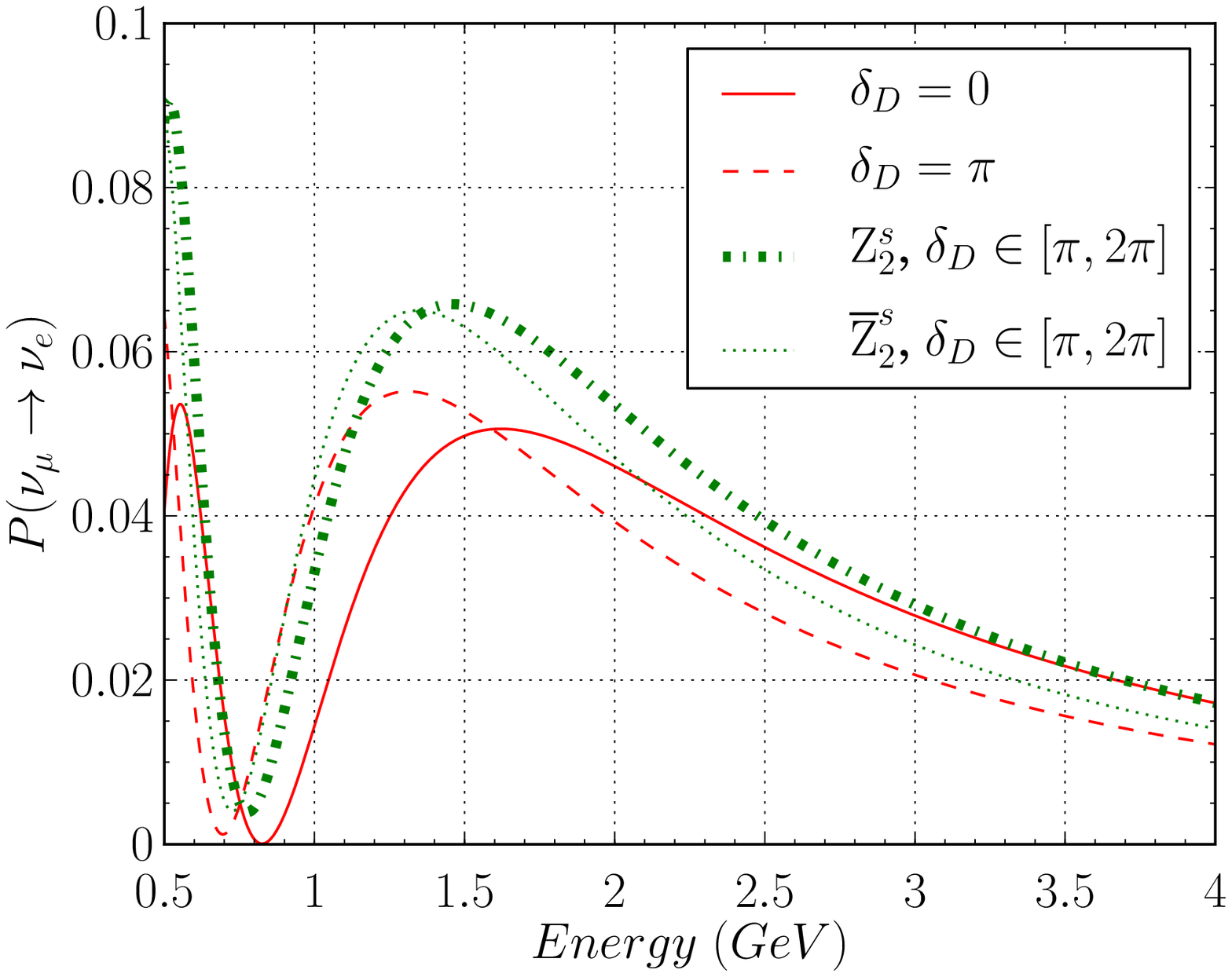}
    \caption{}
    \label{fig:oscillation_NH_NOvA_fogli}
  \end{subfigure}
\end{figure}

\section{Determination of the Mass Hierarchy and the Octant of $\theta_a$}
\label{sec:ellipses}

As has been mentioned previously, a major goal of neutrino oscillation experiments is to determine the mass hierarchy.
If $CP$ was a good symmetry, then there would be
no observable difference between $P(\nu_\mu \to \nu_e)$ and
$P(\overline{\nu}_\mu \to \overline{\nu}_e)$ when the neutrinos are propagating through a vacuum. However,
interestingly enough, the matter effects discussed above emulate the effects of $CP$-violation. Therefore,
there is an observable difference between $P(\nu_\mu \to \nu_e)$ and
$P(\overline{\nu}_\mu \to \overline{\nu}_e)$ even if
$CP$ is a good symmetry. Without the effects of matter the difference between oscillation probabilities for
normal hierarchy versus inverted hierarchy is minimal. Thus it is because of the interactions with matter
that allow for a discernible difference between normal and inverted hierarchy.

\begin{figure}[H]
  \centering
  \begin{subfigure}{0.5\textwidth}
    \centering
    \includegraphics[width=\textwidth]{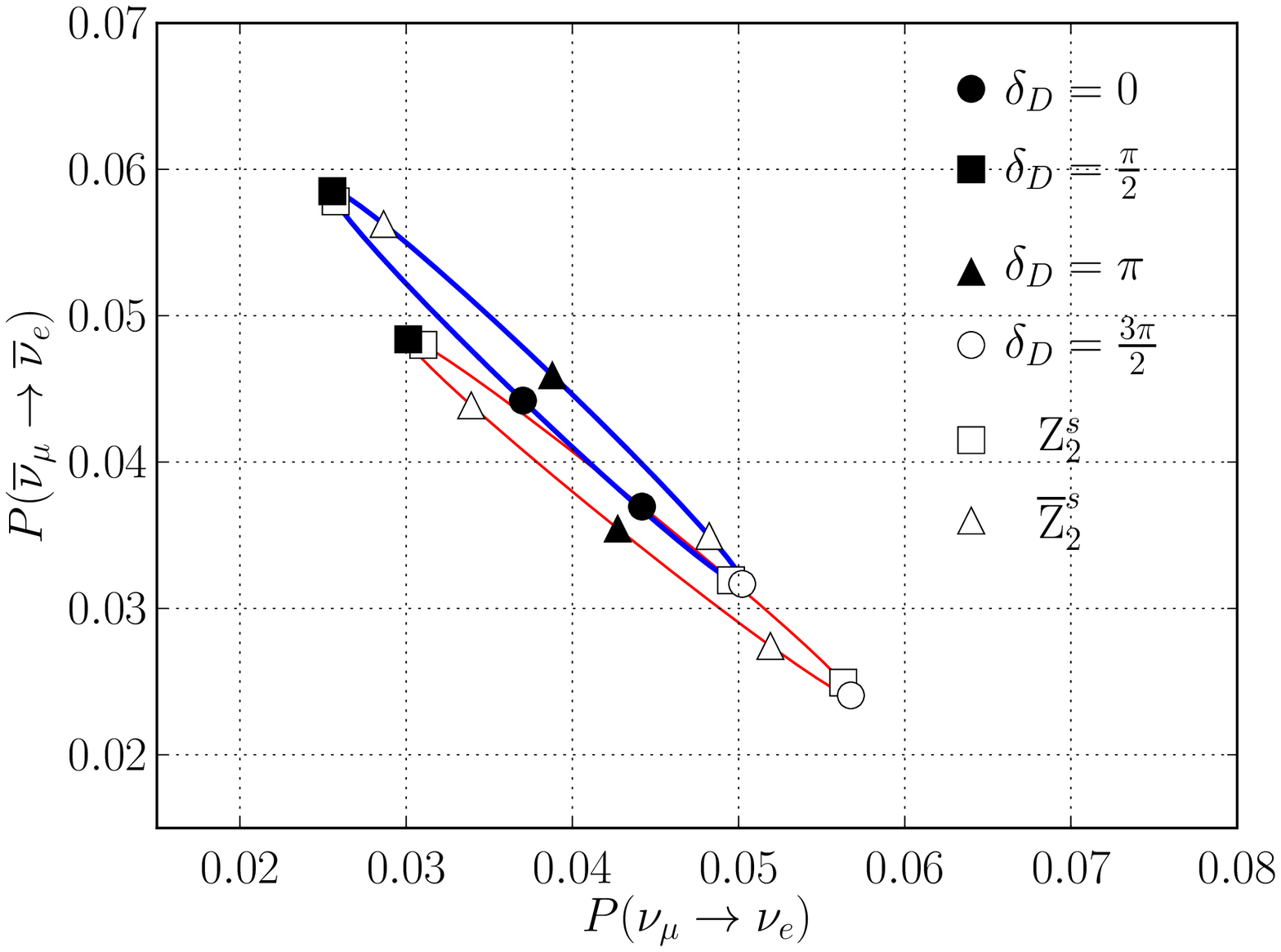}
    \caption{}
    \label{fig:mass_hierarchy_T2K}
  \end{subfigure}
  \caption{Plot of $P(\nu_\mu \to \nu_e)$ vs. $P(\overline{\nu}_\mu \to \overline{\nu}_e)$
           to show the sensitivity of (a) NO$\nu$A (b) T2K to determining the mass hierarchy
           assuming $\theta_a < \frac{\pi}{2}$.
           The red ellipse (lower right corner) corresponds to NH, and the blue ellipse
           (upper left corner) corresponds to IH.}
  \label{fig:mass_hierarchy_fogli}
\end{figure}

It is possible that actual $CP$-violation is substantially cancelled by this
matter induced $CP$-violation. This would be very unfortunate, because it would make the determination
of the $CP$ phase more difficult than expected.
A plot for $P(\nu_\mu \to \nu_e)$ vs. $P(\overline{\nu}_\mu \to \overline{\nu}_e)$ is
shown in Fig. \ref{fig:mass_hierarchy_fogli}
for NO$\nu$A and T2K using the best fits from \cite{Capozzi:2013csa}.
It can be seen that there are many values of
the $CP$ phase that will allow NO$\nu$A to make a serious determination of the true mass hierarchy. This will occur if
$\delta_D \in [\pi, 2 \pi]$ with NH being the true hierarchy, or $\delta_D \in [0,\pi]$ with IH being the true hierarchy.
And since T2K has excluded most of $\delta_D \in [0, \pi]$ at $90\%$ C.L. \cite{Abe:2013hdq}, hopefully the true mass hierarchy is normal.
From Fig. \ref{fig:mass_hierarchy_T2K} it appears that T2K will not be able to determine the mass hierarchy in this manner.

In addition, it may also be possible to determine the

\begin{figure}[H]
  \centering
  \begin{subfigure}{0.5\textwidth}
    \centering
    \includegraphics[width=\textwidth]{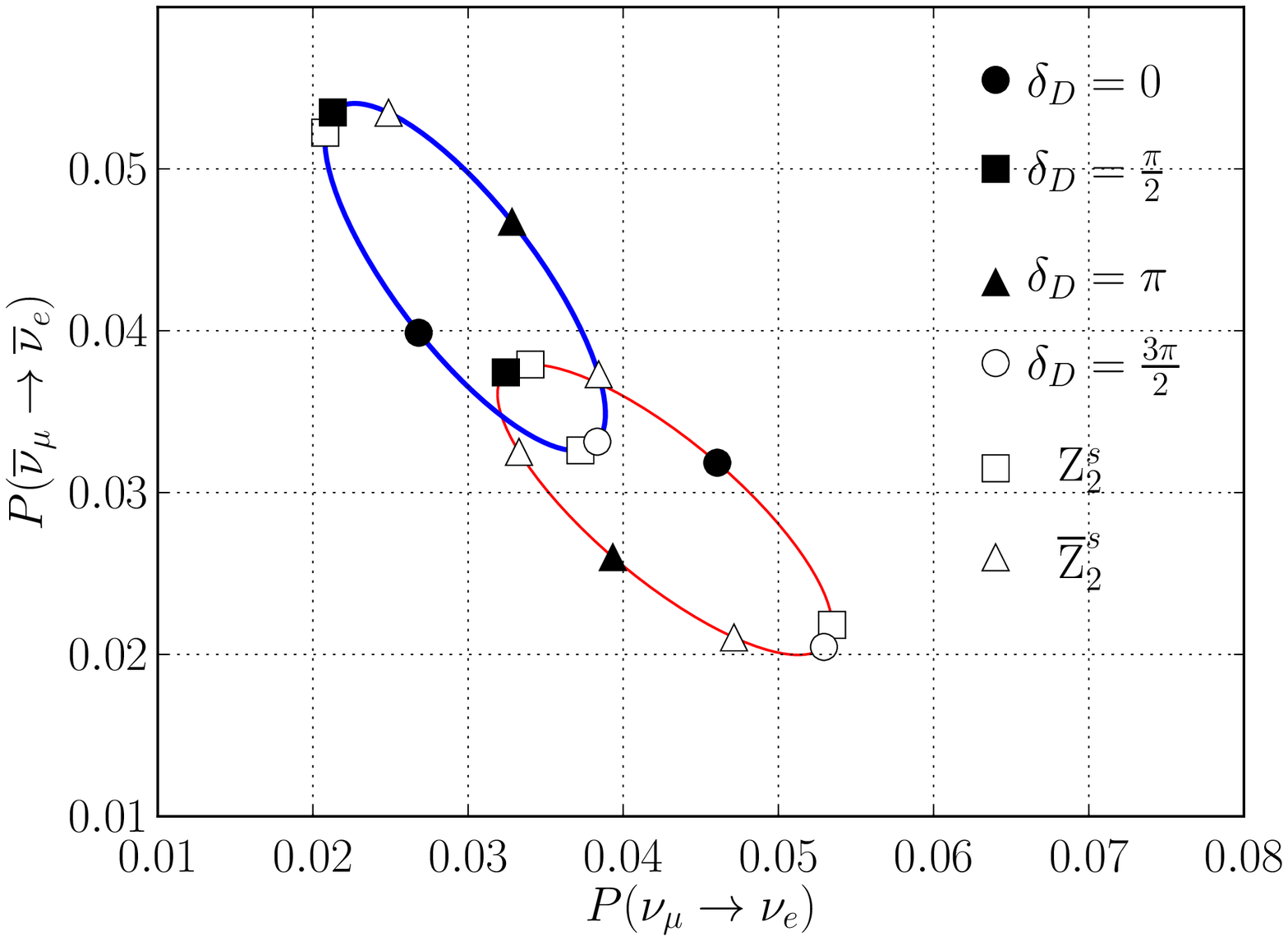}
    \caption{}
    \label{fig:mass_hierarchy_NOvA}
  \end{subfigure}
\end{figure}
\vspace{-40pt}
\begin{figure}[H]
  \begin{subfigure}{0.5\textwidth}
    \centering
    \includegraphics[width=\textwidth]{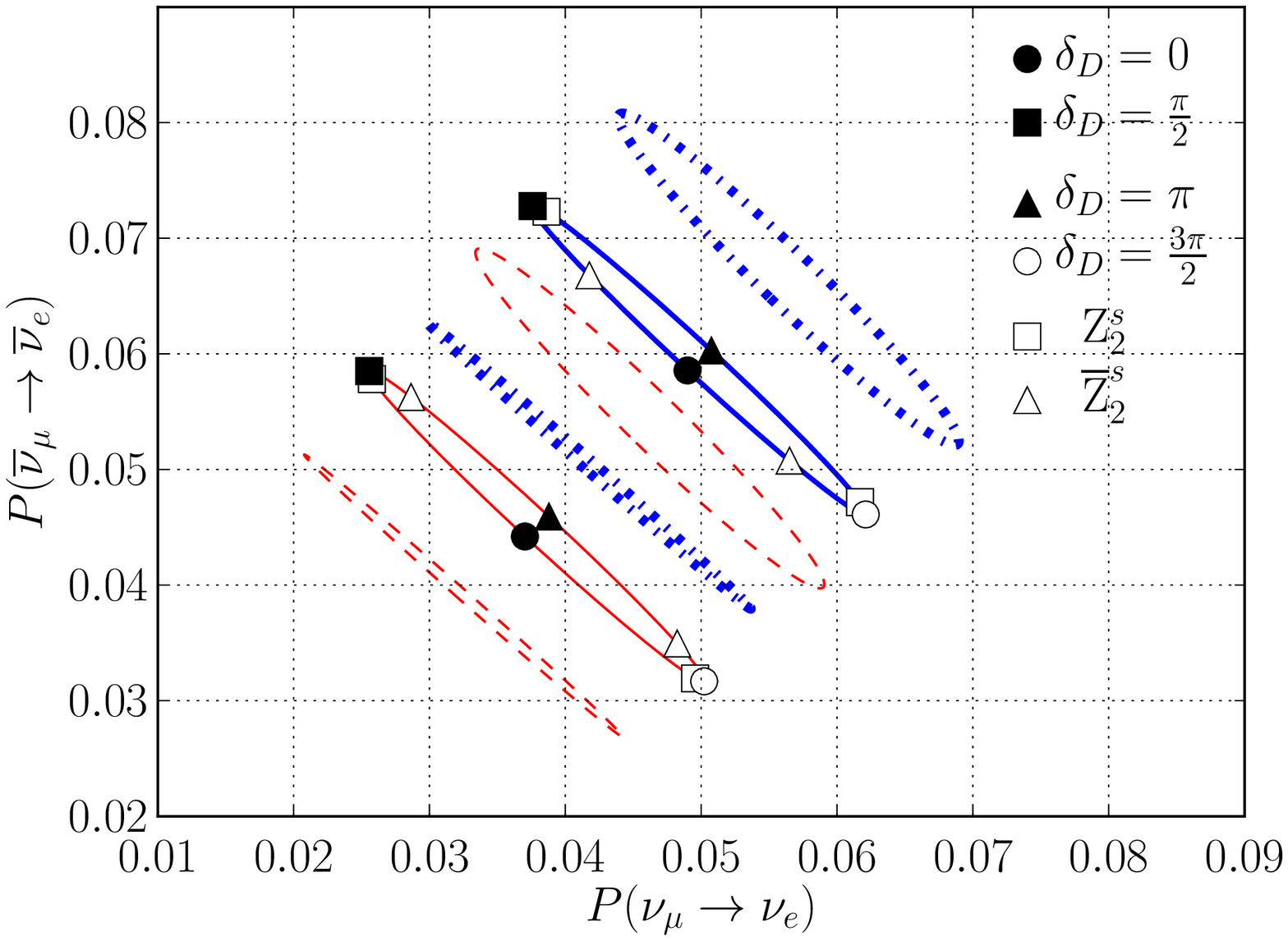}
    \caption{}
    \label{fig:s2a_octant_NH_T2K}
  \end{subfigure}
    
  \caption{Plot of $P(\nu_\mu \to \nu_e)$ vs. $P(\overline{\nu}_\mu \to \overline{\nu}_e)$ to show
           the sensitivity of (a) NO$\nu$A 
           (b) T2K to determining the octant of $\theta_a$ assuming inverted hierarchy. The red ellipses (lower left corner) correspond
           to $\theta_a < \frac{\pi}{4}$, while the blue ellipses (upper right corner) correspond to
           $\theta_a > \frac{\pi}{4}$. The dashed ellipses correspond to the $\pm 1 \sigma$ values.}
  \label{fig:s2a_octant_fogli_NH}
\end{figure}

\noindent
octant  of $\theta_a$ from similar plots.
These are shown in Fig. \ref{fig:s2a_octant_fogli_NH}. It appears that every value of the $CP$ phase could at least give some 
indication of the true octant of $\theta_a$, but
the best values would be $\delta_D = 0$ for the lower octant and $\delta_D = \pi$
for the higher octant.

The ellipses were created by using (\ref{eq:prob_mu_to_e}) with the matter effect modifications of (\ref{eq:eff_angles}).
for all possible values of $\delta_D$ (i.e.
$\delta_D$ $\in$ $[0, 2 \pi]$). The $\Box$ and the $\bigtriangleup$ symbols correspond to the predicted values for
$\delta_D$, based on $\mathbb{Z}^s_2$ and $\overline{\mathbb{Z}}^s_2$, respectively. The predicted values
are determined by using the best fits from \cite{Capozzi:2013csa} in (\ref{eq:cD}).

\begin{figure}[H]
  \centering
  \begin{subfigure}{0.5\textwidth}
    \centering
    \includegraphics[width=\textwidth]{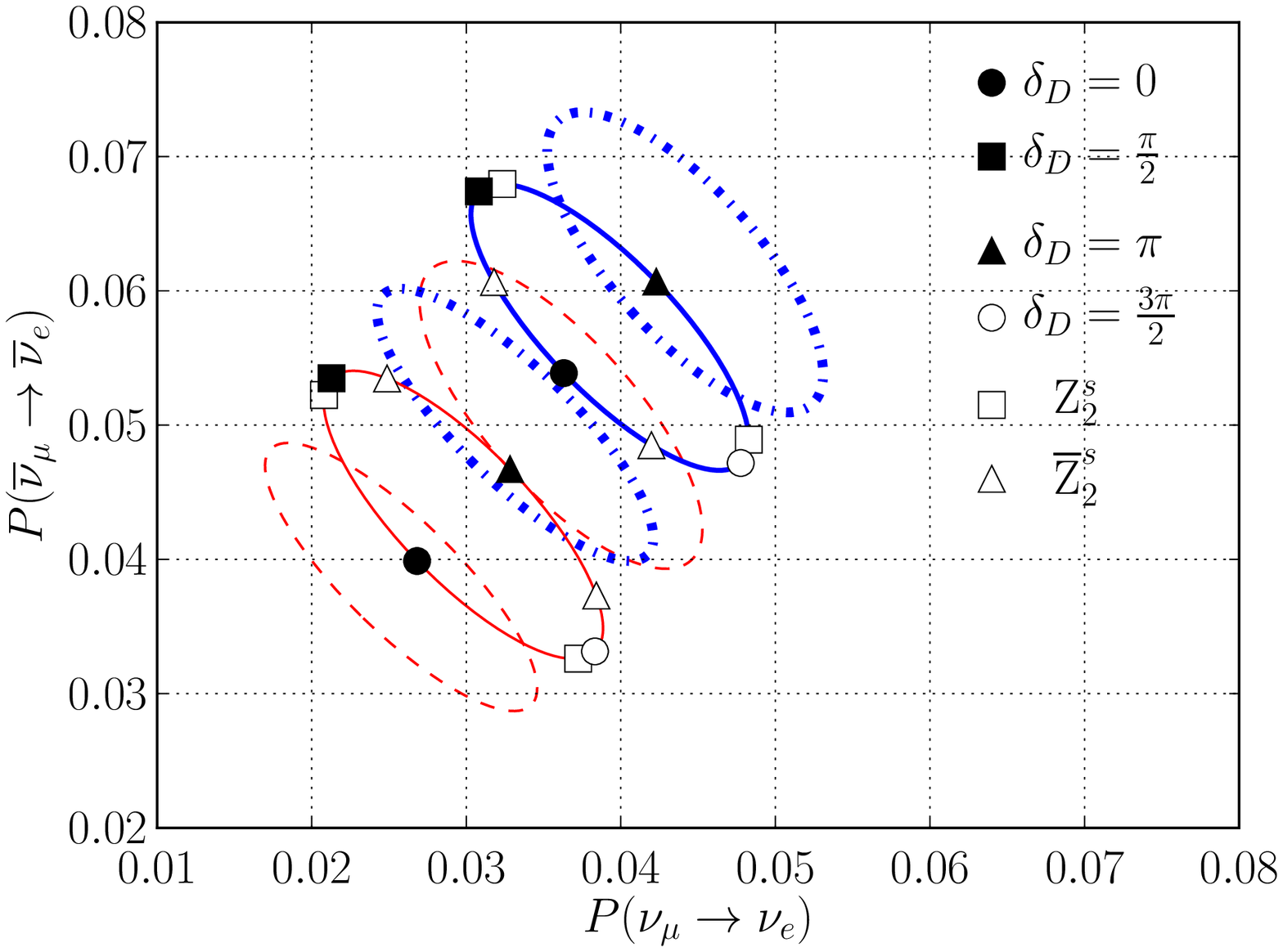}
    \caption{}
    \label{fig:s2a_octant_NH_NOvA}
  \end{subfigure}
\end{figure}
\vspace{-50pt}

\section{Effective Masses and Mixing Angles in Matter}

\begin{figure}[H]
  \begin{subfigure}{0.5\textwidth}
    \centering
    \includegraphics[width=\textwidth]{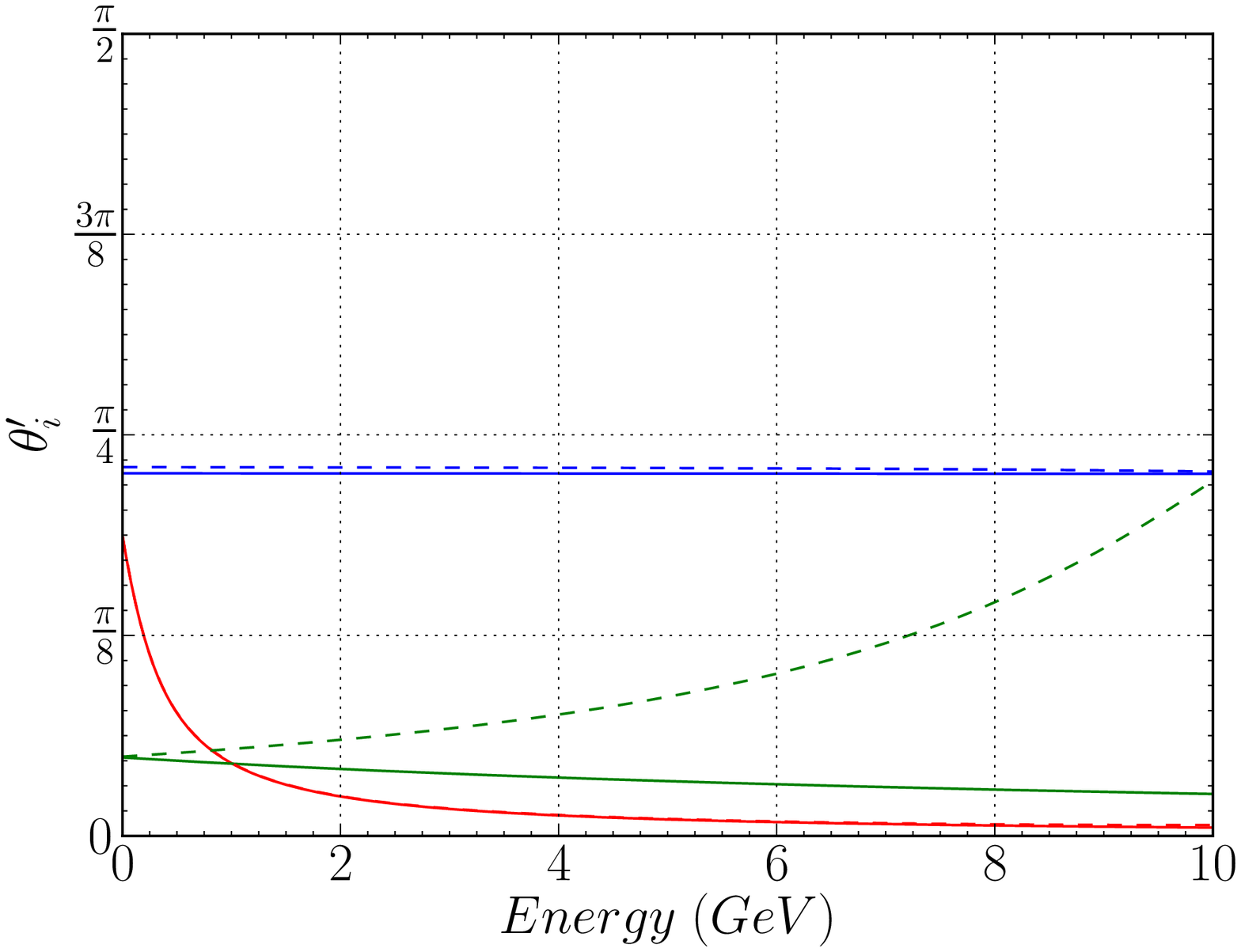}
    \caption{}
    \label{fig:sin2_eff_Z_anti_NOvA}
  \end{subfigure}
  \caption{Plots of the effective mixing angles as a function of energy using data from \cite{Capozzi:2013csa}.
           For (a) NO$\nu$A and neutrinos, (b) NO$\nu$A and anti-neutrinos.
           Key: red $= \theta^\prime_s$, blue $= \theta^\prime_a$, and green $= \theta^\prime_r$. The solid lines are for
           normal hierarchy, and the dashed lines are for inverted hierarchy. On the plots where the dashed line
           is not visible, it's because the solid line is on top of it.}
  \label{fig:sin2_eff}
\end{figure}

\begin{figure}[H]
  \centering
  \begin{subfigure}{0.47\textwidth}
    \centering
    \includegraphics[width=\textwidth]{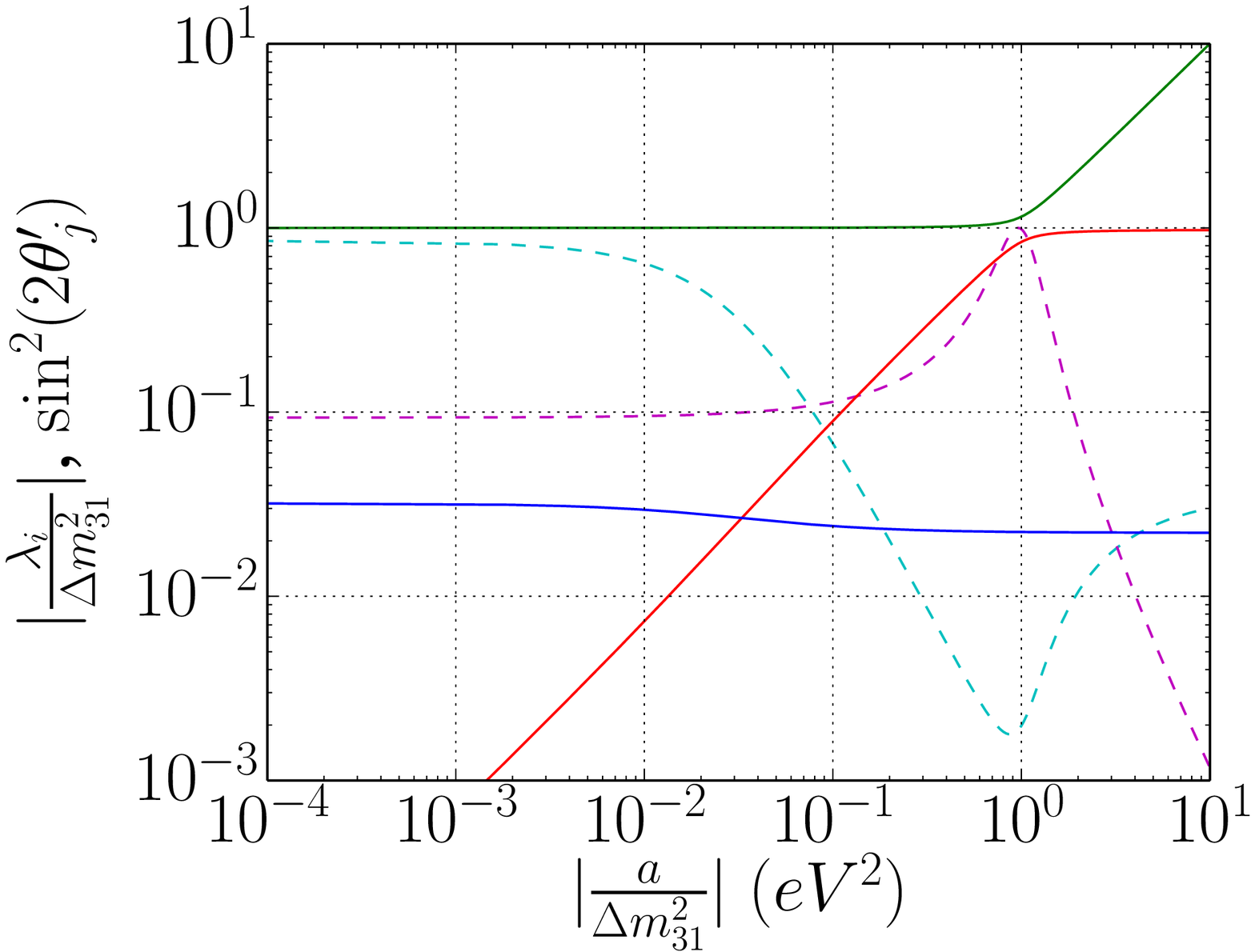}
    \caption{}
    \label{fig:sin2_2_masses_eff_anti_NOvA_IH}
  \end{subfigure}
  \caption{Effective masses for NO$\nu$A. The major focus of these
           plots should be on the solar and atmospheric resonances corresponding
           to a level repulsion.
           (a) $\nu$ and NH, (b) $\nu$ and IH,
           (c) $\overline{\nu}$ and NH, and (d) $\overline{\nu}$ and IH.
           Key: red $= \lambda_1$, blue $= \lambda_2$, green $= \lambda_3$,
                cyan-dashed $= \sin^2 (2 \theta_s)$, and magenta-dashed $= \sin^2 (2 \theta_r)$.}
  \label{fig:sin2_2_masses_eff_NOvA}
\end{figure}

The values of the effective mixing angles are plotted in Fig. \ref{fig:sin2_eff}, and the mass eigenvalues in Fig. \ref{fig:sin2_2_masses_eff_NOvA},
as functions of energy using the matter density for the NO$\nu$A experiment. The plots for T2K are excluded here,
because they don't differ much from the ones for NO$\nu$A.
Also, these particular plots consider  $\delta_D = 0$, because the results depend very little on the $CP$ phase.
These have
\begin{figure}[H]
  \centering
  \begin{subfigure}{0.5\textwidth}
    \centering
    \includegraphics[width=\textwidth]{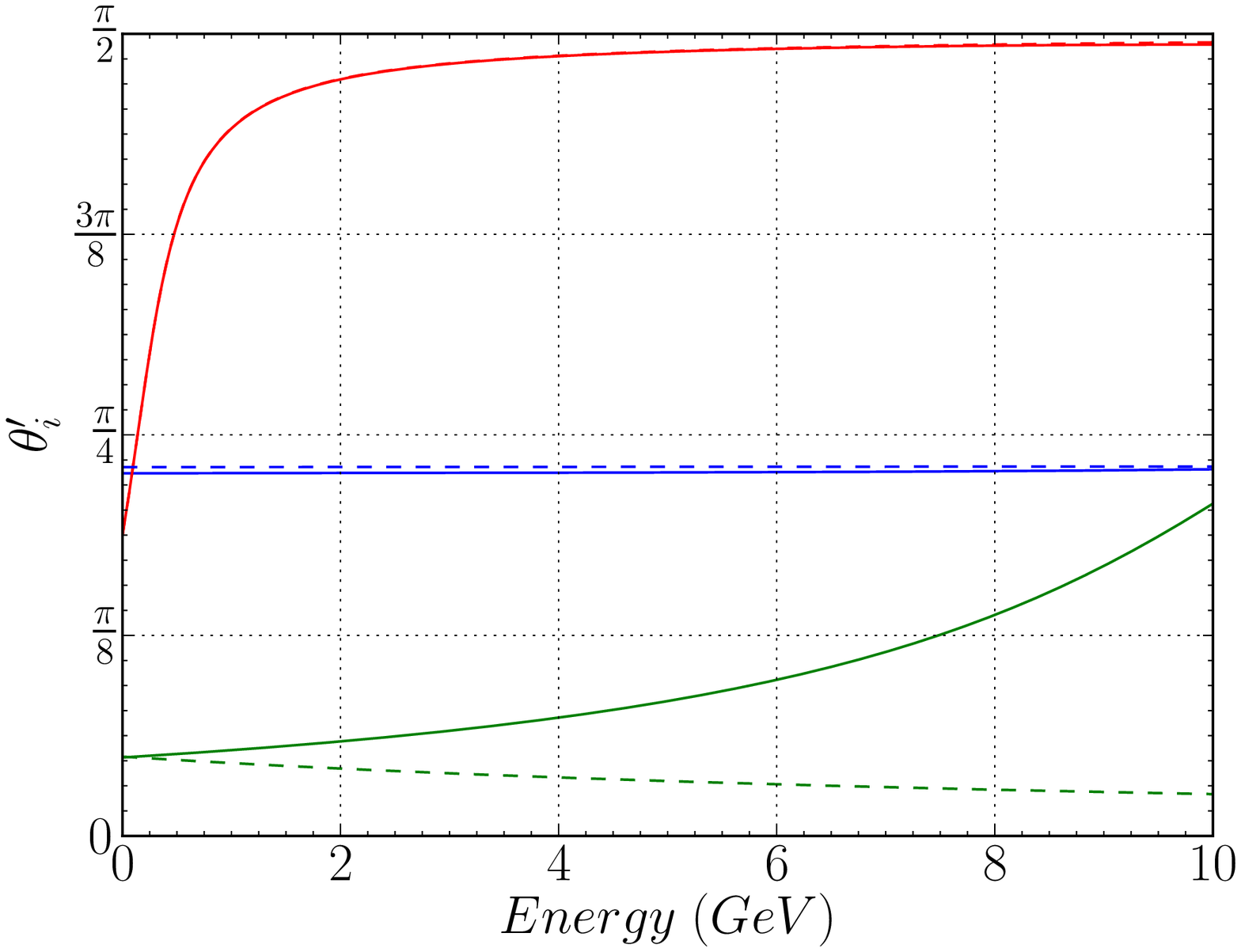}
    \caption{}
    \label{fig:sin2_eff_Z_NOvA}
  \end{subfigure}
\end{figure}
\vspace{-26pt}
\noindent
been plotted by numerically diagonalizing the Hamiltonian.
It's assumed the diagonalization matrix will have
the same form as the
standard parameterization of the PMNS mixing matrix.

The approximation introduced in Sec. \ref{sec:matter} implies that the $CP$ phase
and $\theta^\prime_a$ do not vary much, if at all, due to interactions with matter (which can be observed in
Fig. \ref{fig:sin2_eff}). It also implies certain
characteristics of the variations of the other two mixing angles. 
From (\ref{eq:theta_s_approx}),
$\theta^\prime_s$ should be independent of the mass hierarchy, and taking the limit
$a \to \infty$, then $\theta^\prime_s \to \frac{\pi}{2}$ ($0$) for $\nu$ ($\overline{\nu}$). This behavior is easily observed
in Fig. \ref{fig:sin2_eff}.
From (\ref{eq:theta_r_approx}), $\theta^\prime_r$ should have similar asymptotic behavior as $\theta^\prime_s$ for normal 

\begin{figure}[H]
  \centering
  \begin{subfigure}{0.5\textwidth}
    \centering
    \includegraphics[width=\textwidth]{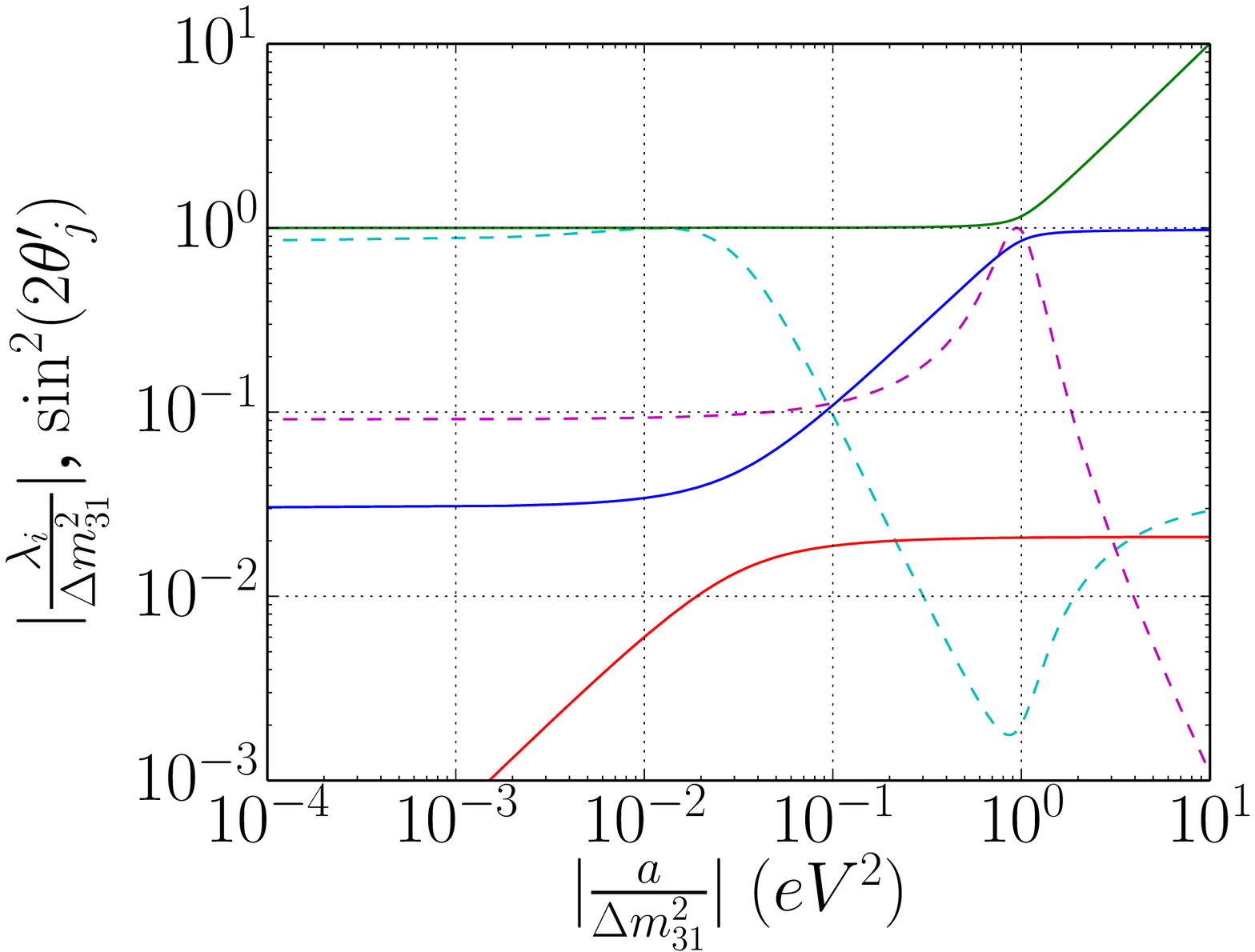}
    \caption{}
    \label{fig:sin2_2_masses_eff_NOvA_NH}
  \end{subfigure}
\end{figure}

\noindent
hierarchy, while
it should reverse its behavior for inverted hierarchy.
These features are approximately shown in Fig. \ref{fig:sin2_eff}, but
at the energies shown, $\theta^\prime_r$ is not able to approach its asymptotic limit. Therefore, these results
appear to agree with the approximation in \cite{Agarwalla:2013tza}.

The effective neutrino masses are found from multiplying the eigenvalues of the Hamiltonian by $2 E$. These plots are shown
in Fig. \ref{fig:sin2_2_masses_eff_NOvA}
for NO$\nu$A.
There are some interesting characteristics of these plots. The first and most obvious
are two resonances referred to as the solar resonance and the atmospheric resonance which
represent the condition for maximal oscillation probability. This phenomenon was first
understood with the introduction of the MSW effect \cite{Wolfenstein:1977ue,Mikheev:1986wj}.
The first peak of $\sin^2 (2 \theta^\prime_s)$ is the solar resonance
and corresponds to an approach of $|\lambda_1|$ and $|\lambda_2|$ followed by a repulsion. The
first peak of $\sin^2 (2 \theta^\prime_r)$ is the atmospheric resonance and corresponds to an approach
of $|\lambda_2|$ and $|\lambda_3|$ followed by a repulsion. If the absolute value of the 
mass eigenvalues cross, then no resonance can be seen there. If we don't
take the absolute value of the mass eigenvalues, then they will never
cross each other. This is a 
wonderful example of level
repulsion in quantum mechanics.
For more details on
these resonances, including a derivation of the resonance condition, see
\cite{Wolfenstein:1977ue,Mikheev:1986wj,Smirnov:2004zv,Friedland:2001xk,PhysRevD.86.010001}.

\section{Conclusions}

Predicted distributions for $\delta_D$, $J_\nu$, and $\theta_a$ were updated using the residual symmetries
$\mathbb{Z}^s_2$ and $\overline{\mathbb{Z}}^s_2$. It was found that the greater uncertainty in the octant
of $\theta_a$ for IH shown in \cite{Capozzi:2013csa} forced the distributions of $\delta_D$ for IH to 
have nearly equal contributions on either side of $\delta_D = -90^\circ$. This had no significant effect
on the distribution for $J_\nu$ and the prediction for $J_\nu$ has improved.

By including the effects of matter into the oscillation probabilities, it was shown in Sec. \ref{sec:ellipses} how NO$\nu$A stands
\begin{figure}[H]
  \centering
  \begin{subfigure}{0.48\textwidth}
    \centering
    \includegraphics[width=\textwidth]{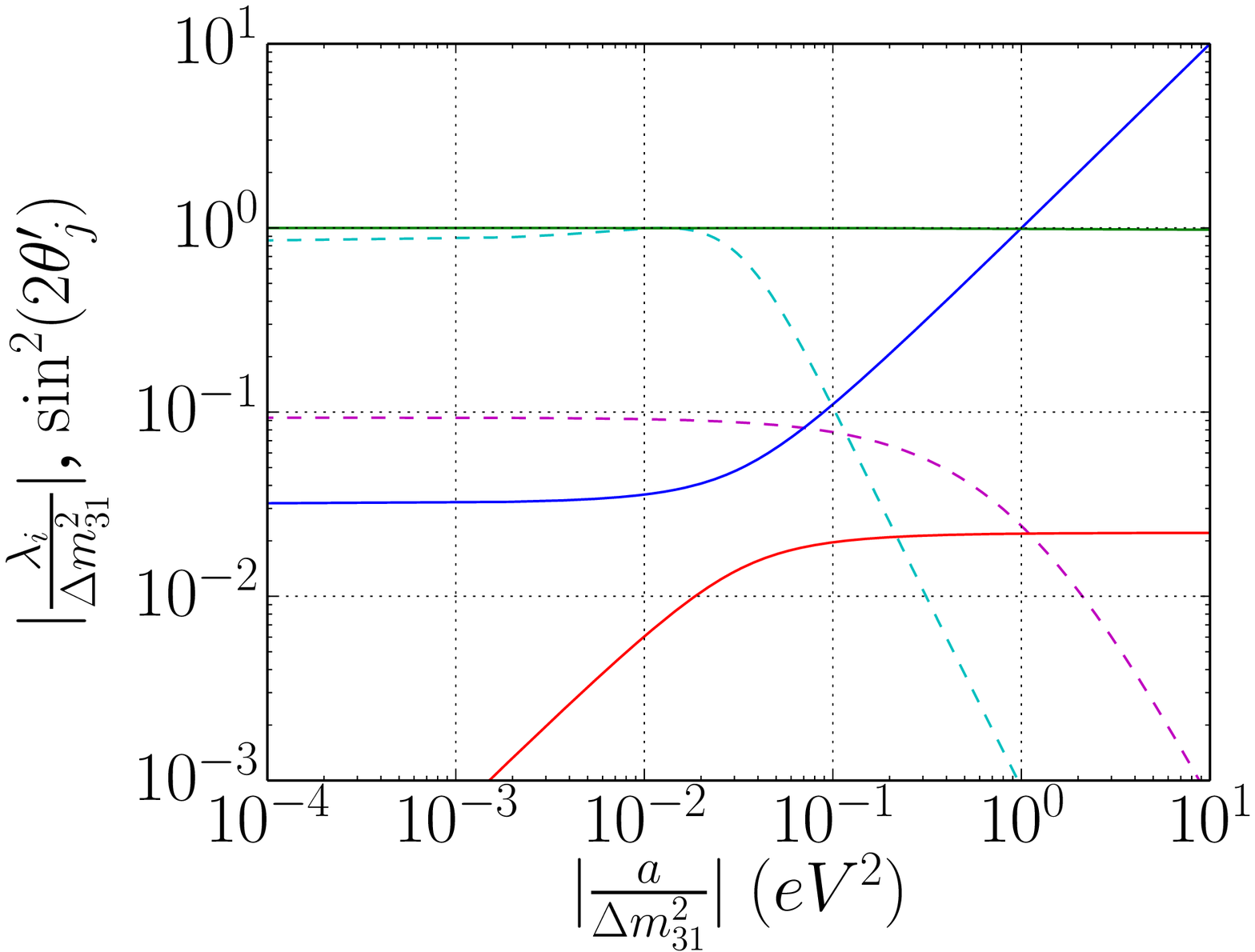}
    \caption{}
    \label{fig:sin2_2_masses_eff_NOvA_IH}
  \end{subfigure}
\end{figure}
\vspace{-26pt}
\begin{figure}[H]
  \centering
  \begin{subfigure}{0.47\textwidth}
    \centering
    \includegraphics[width=\textwidth]{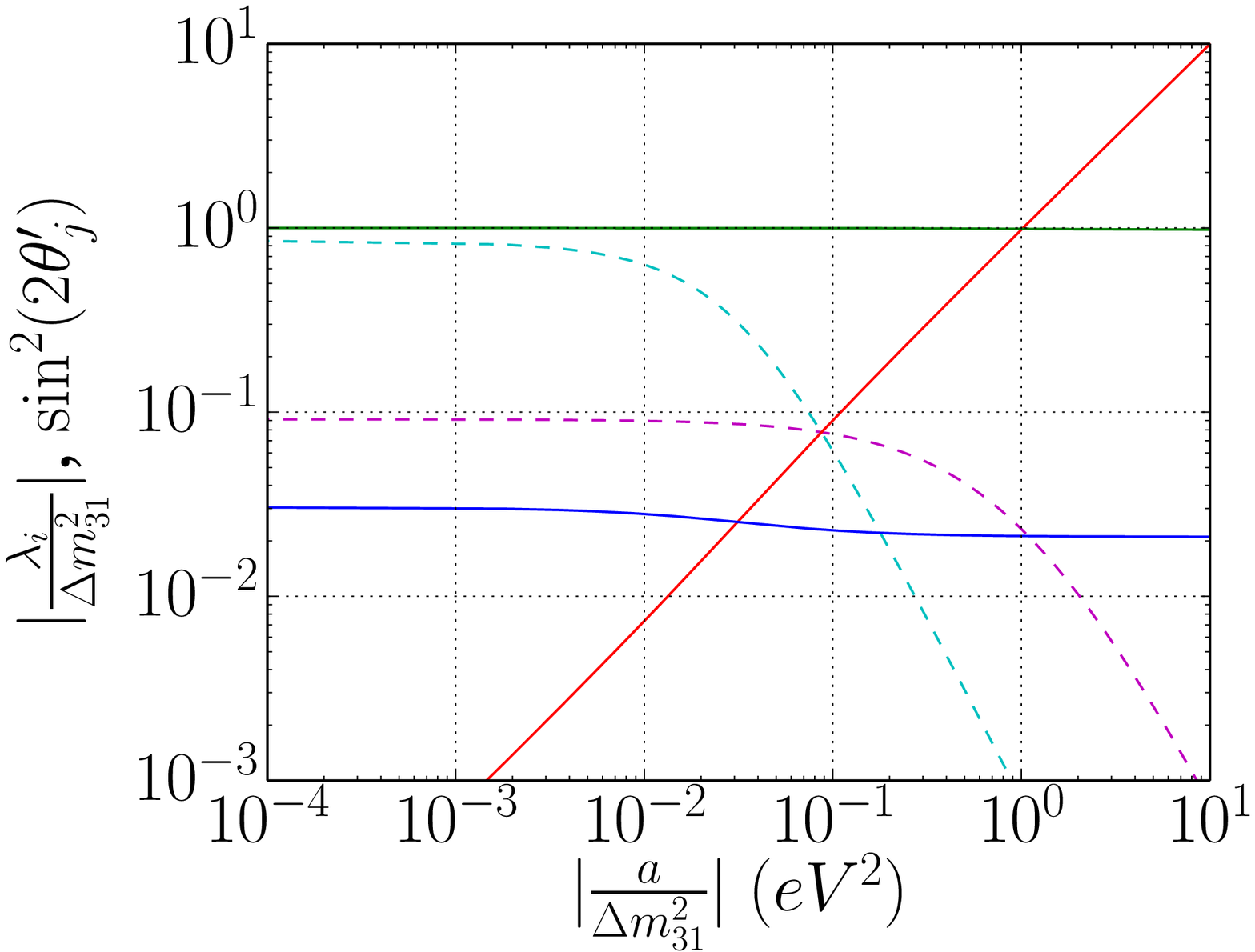}
    \caption{}
    \label{fig:sin2_2_masses_eff_anti_NOvA_NH}
  \end{subfigure}
\end{figure}
\vspace{-26pt}

\noindent
a good chance to determine the mass hierarchy if $\delta_D \in [\pi, 2 \pi]$ and the true hierarchy is normal, or if
$\delta_D \in [0, \pi]$ and the true hierarchy is inverted.
It was also shown that both NO$\nu$A and T2K may be capable
of nailing down the octant of $\theta_a$.

The effects of matter were also shown to give rise to two resonances: the solar resonance and the atmospheric resonance.
This behavior can be seen to agree with the approximation used throughout this work \cite{Agarwalla:2013tza}.

\section{Acknowledgments}

The authors would like to thank Shao-Feng Ge for the contributions in the early
stages of this work. W.W.R. would also like to thank Kendall Mahn for some helpful conversations.
All plots in this paper were produced using matplotlib \cite{Hunter:2007}.
W.W.R. was supported in part by the National Science Foundation under Grant No. PHY-1068020.
D.A.D. was supported in part by the U.S. Department of Energy under Award No. DE-FG02-12ER41830.
\vspace{-10pt}
\begin{figure}[H]
  \centering
  \begin{subfigure}{0.45\textwidth}
    \centering
    \includegraphics[width=\textwidth]{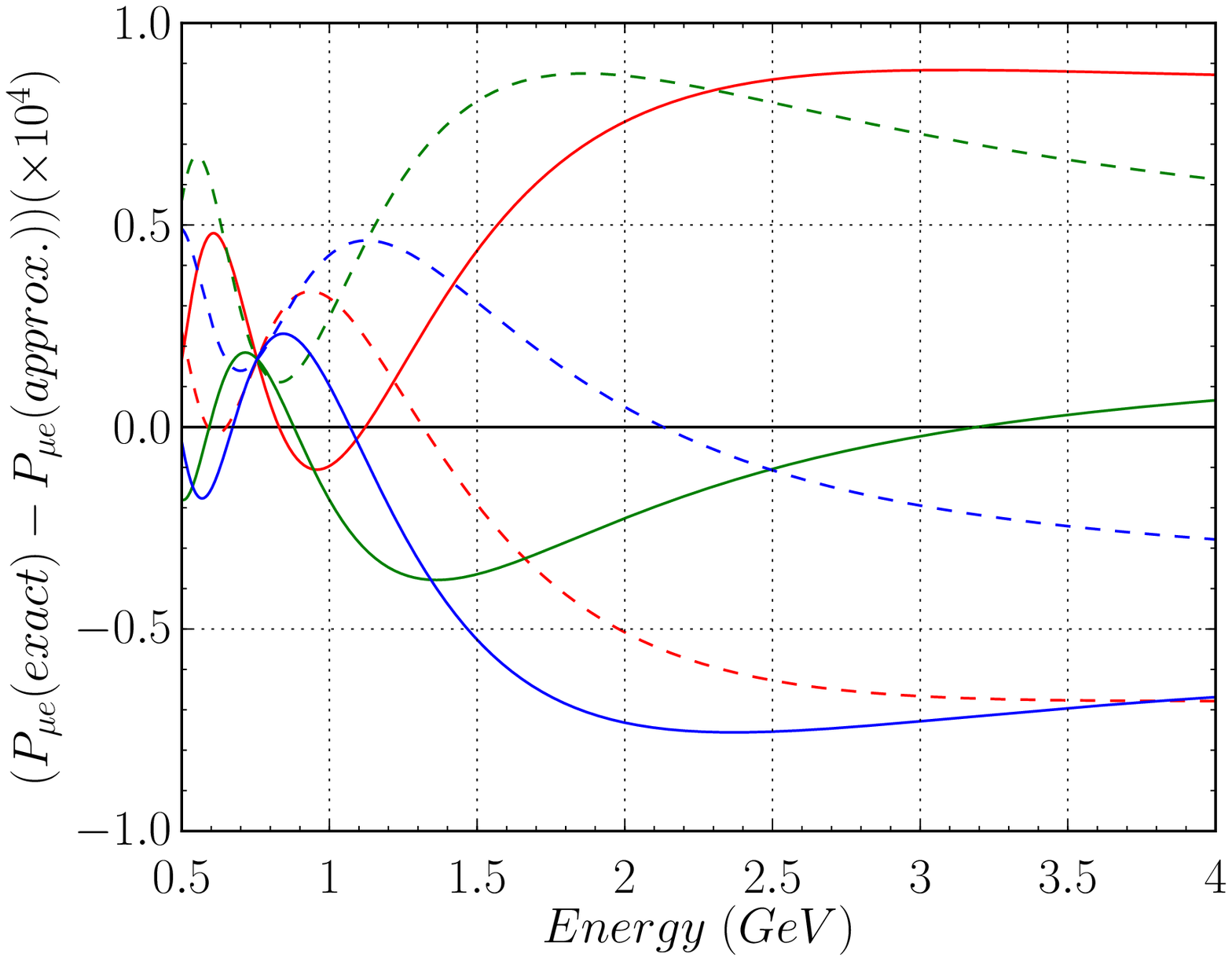}
    \caption{}
    \label{fig:compare_NOvA}
  \end{subfigure}
\end{figure}


\appendix

\section{Comparison with Solving for Matter Effects Exactly}

Here a comparison is made between the approximation used \cite{Agarwalla:2013tza} and
exact results found from numerically diagonalizing the Hamiltonian. Each plot for 
$P(\nu_\mu \to \nu_e)$ and $P(\overline{\nu}_\mu \to \overline{\nu}_e)$ above has been redone without
any approximation. The plots below show the difference between these two methods. It's clear that
the approximation is indeed very good, with a maximum difference around $0.0001$.
\vspace{-10pt}
\begin{figure}[H]
  \centering
  \begin{subfigure}{0.45\textwidth}
    \centering
    \includegraphics[width=\textwidth]{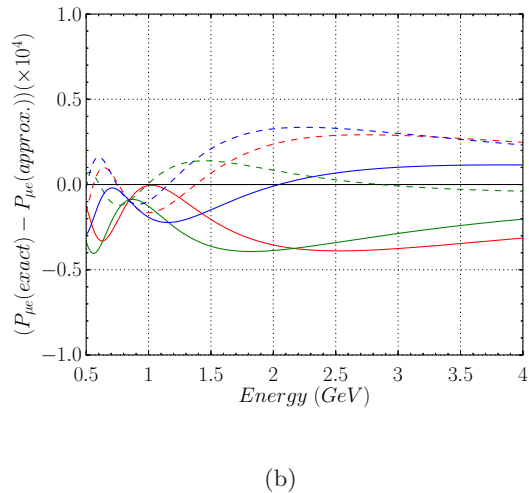}
    \label{fig:compare_anti_NOvA}
    \caption{}
  \end{subfigure}
  \caption{Comparison between the exact results and
           the approximation used throughout the paper.
           For (a) NO$\nu$A and neutrinos, (b) NO$\nu$A and anti-neutrinos.
           Key: red $= (\delta_D = 0)$, red-dash $= (\delta_D = \pi)$, green $= \mathbb{Z}^s_2, \delta_D \in [0, \pi]$,
                green-dash: $= \mathbb{Z}^s_2, \delta_D \in [\pi, 2 \pi]$, blue: $= \overline{\mathbb{Z}}^s_2, \delta_D \in [0, \pi]$,
                blue-dash: $= \overline{\mathbb{Z}}^s_2, \delta_D \in [\pi, 2 \pi]$.}
\end{figure}

\bibliography{bib}

\end{document}